\documentclass[aps,prl,reprint,superscriptaddress,showpacs]{revtex4-2}
\usepackage{amsmath,amsfonts}
\usepackage{xcolor}
\usepackage[colorlinks,linkcolor=blue, urlcolor=blue, anchorcolor=blue, citecolor=blue]{hyperref}
\usepackage[T1]{fontenc}
\usepackage{newtxtext,newtxmath}

\usepackage{dcolumn}%
\usepackage{graphicx}
\usepackage{bm}

\usepackage{amssymb}
\usepackage{setspace}
\usepackage{romannum}

\DeclareMathAlphabet{\mathcal}{OMS}{cmsy}{m}{n}  

\begin{document}
	

	\title{Enhanced Quantum Metrology via Saddle-Point Scrambling in Phase Space}
	
	\author{Lei Shao}
	\affiliation{Graduate School of China Academy of Engineering Physics, Beijing 100193, China}
	\author{Hai-Jun Xing}
	\affiliation{Center for Quantum Sciences and School of Physics, Northeast Normal University, Changchun 130024, China}
	\author{Libin Fu}
    \email{lbfu@gscaep.ac.cn}
   	\affiliation{Graduate School of China Academy of Engineering Physics, Beijing 100193, China}



	\begin{abstract}
Nonlinear effects are widely utilized in quantum metrology to enhance measurement precision by leveraging complex dynamical behaviors. Here, we propose a quantum scrambling-integrated scheme that optimizes the selection of initial states and parameter encoding by exploiting the nonlinear dynamical trajectories and their geometric properties in phase space. We consider a two-mode Bose-Einstein condensate as a typical example. Within a certain parameter range, using uncorrelated spin coherent states located at saddle points and separatrices enables the measurement precision to reach the Heisenberg scaling. An analytical expression for the quantum Fisher information under long-time evolution is provided, and our analysis further reveals that the enhancement mechanism is governed by the sensitivity of the energy spectrum to parameter variations. Compared to the criticality-enhanced protocol that requires parameters to approach the critical point, our scheme offers a distinct advantage in its applicability over a broader range of parameters. Moreover, this scheme can also be applied to models that include one-axis and two-axis countertwisting interactions, and chaotic dynamics. This work offers a new perspective for advancing the application of nonlinear dynamics to enhance the parameter estimation precision in quantum systems.

	\end{abstract}
	
	
	\maketitle
	

	\textit{Introduction}.--- Exploring valuable nonclassical resources to enhance the precision of parameter estimation is a crucial task in quantum metrology.  Quantum properties such as entanglement and squeezing have garnered significant attention due to their potential applications in high-precision measurements, including gravitational wave detection~\cite{PhysRevA.88.041802,RevModPhys.86.121}, biological sensing~\cite{TAYLOR20161,NASR20091154}, and inertial navigation~\cite{Giovannetti2001}. By leveraging these nonclassical resources, measurement precision can surpass the standard quantum limit (SQL)~\cite{Sørensen2001,Orzel2001,Gross2010,PhysRevA.64.013616,PhysRevA.67.013607,PhysRevA.72.033612,Riedel2010,Estève2008}. To effectively enhance measurement precision, a prevalent approach is to generate nonclassical states using nonlinear interactions, leading current research to focus on preparing states suitable for metrology within a limited time~\cite{PhysRevA.90.062132,PhysRevA.92.013623,PhysRevLett.117.013001,PRXQuantum.4.020314}. A promising avenue for achieving this goal lies in leveraging semi-classical dynamics within quantum metrology. For instance, by analysing classical equations of motion and the evolution phase-space trajectory, the state preparation process in collective spin systems can be optimized to generate metrologically useful states within a shorter timescale~\cite{PRXQuantum.4.020314,PhysRevA.92.013623}. However, nonclassical states are fragile, difficult to stabilize, and susceptible to decoherence. One approach to addressing this issue is to focus on parameter encoding through nonlinear dynamics~\cite{Fiderer2018}.
	
	 In addition, quantum scrambling~\cite{Sekino2008,Lewis-Swan2019,PhysRevA.94.040302,PhysRevLett.124.140602,PhysRevA.103.033304,PRXQuantum.5.010201,doi:10.1126/science.adg9500,doi:10.1073/pnas.2217031120,39bt-37yl} which can be induced by nonlinear interactions and describes the local information spreading across the system's multiple degrees of freedom, has also been applied to quantum metrology. A recent experiment in Ref.~\cite{doi:10.1126/science.adg9500} has demonstrated synchronized exponential growth of metrological gain and out-of-time-order correlators (OTOCs) through information scrambling near bistable points in phase space, further validating the feasibility of integrating quantum-classical correspondence into metrological protocols.  To harness quantum scrambling as a metrological resource, clarifying its precision enhancement mechanism is critical. Nevertheless,  research on this aspect is currently lacking. Parallel to these developments,  criticality-enhanced metrology~\cite{PhysRevA.78.042105,PhysRevA.78.042106,PhysRevX.8.021022,Yang:19,PRXQuantum.3.010354,PhysRevLett.125.240506,PhysRevLett.124.120504,PhysRevLett.126.010502,PhysRevLett.132.060801,Garbe_2022}, which utilizes quantum phase transitions as valuable resources, has also become a prominent research focus. One typical method involves using a sudden quench to encode the parameter information around the critical point~\cite{PhysRevLett.126.010502,Garbe_2022,PhysRevA.110.012413,Zhu2023}, and it commonly requires prior knowledge and precise control of phase transition parameters. This strict requirement, which confines viable parameters to a narrow region and reduces practicality, motivates the need for alternative strategies. The method to address this stringent condition is by exploiting the nonlinear interactions inherent in the criticality-enhanced protocol. Through analyzing classical dynamics, the selection of initial states can be optimized, thereby expanding the range of feasible parameters.
	
	In this Letter, we introduce a framework termed the quantum scrambling-integrated (QSI) scheme to optimize the parameter encoding in nonlinear dynamics. By exploiting the geometric properties of classical trajectories and saddle-point scrambling, our protocol optimizes the selection of initial states. Specifically, preparing initial states at saddle points and along separatrices provides a metrological advantage, enabling precision to reach the Heisenberg scaling. Here, we use quantum Fisher information (QFI) to quantify the parameter sensitivity,  while introducing the concept of gain factor to further elucidate the intrinsic mechanism underlying the metrological advantage associated with the saddle points and separatrices. Our results show that initial states on the separatrices exhibit superior performance and have potential for metrological applications. Compared to the quench method, our approach broadens the accessible parameter region,  making the protocol more practical and feasible. Furthermore, to assess the scheme’s viability in realistic scenarios, we evaluate its performance using the classical Fisher information (CFI)  while accounting for the impact of decoherence~\cite{PhysRevLett.72.3439,Xing_2024,PhysRevResearch.5.013185,PhysRevA.92.023603,PhysRevLett.105.020401,PhysRevA.110.032612}.

\begin{figure}[tbp]
		\includegraphics[width=\linewidth]{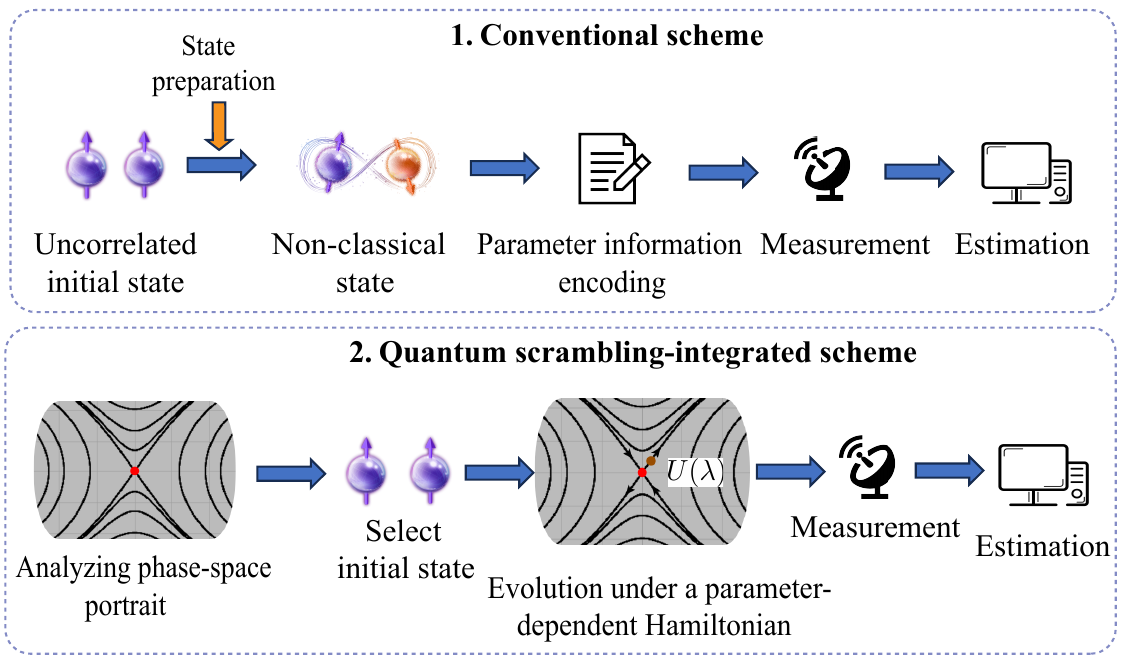}
		\caption{\label{fig:1} Schematic of the conventional and quantum scrambling-integrated schemes. In conventional schemes, an uncorrelated initial state is first prepared into a nonclassical state via a state preparation process, followed by parameter-dependent linear interactions for parameter encoding, and finally completed with measurements and estimation. In contrast, for the quantum scrambling-integrated scheme, the first step involves selecting appropriate points on the phase-space trajectory as the initial state, followed by the evolution under a parameter-dependent Hamiltonian, ultimately leading to measurement and estimation in a similar manner.
		}
	\end{figure}

	\textit{Quantum scrambling-integrated scheme}.--- The conventional parameter estimation scheme can be divided into four steps: (\romannumeral1) preparation of the probe state; (\romannumeral2) encoding of the parameter information; (\romannumeral3) measurements of the observables; (\romannumeral4) estimation of the data (as shown in Fig.\ref{fig:1})~\cite{RevModPhys.90.035005,liu2020quantum,Escher2011}. In the state preparation process, a widely used method for generating nonclassical states is to obtain the desired state through the dynamics under nonlinear interactions~\cite{MA201189,PhysRevA.90.062132,PhysRevA.92.013623,PRXQuantum.4.020314,PhysRevLett.102.100401,PhysRevA.109.052618}. After state preparation, parameter information encoding requires a subsequent adjustment to the Hamiltonian (e.g., applying an external magnetic field $H=- \gamma B S_{z}$~\cite{MA201189,PhysRevA.79.062107,PhysRevA.90.062132}). This necessitates preserving the prepared states during the Hamiltonian modifications, which may increase the experimental implementation complexity. 
	
	In the QSI scheme, we optimize the selection of the initial state by analyzing the phase-space portrait of the Hamiltonian containing both linear parameter-dependent and nonlinear terms, as well as potential saddle-point scrambling, ultimately achieving the aim of enhanced precision. 
	Specifically, this scheme can be divided into the following steps: (\romannumeral1) compute the phase-space trajectories for the system being studied via the mean-field approximation. (\romannumeral2) select an appropriate uncorrelated spin coherent state $\left|\theta_{0},\phi_{0}\right\rangle =\otimes_{n=1}^{N}\left[\cos\left(\theta_{0}/2\right)\left|\uparrow^{\left(n\right)}\right\rangle +\sin\left(\theta_{0}/2\right)e^{i\phi_{0}}\left|\downarrow^{\left(n\right)}\right\rangle \right] $ as the initial state based on the geometric properties of the classical trajectories, where $\theta_{0}$ and $\phi_{0}$ are the polar angle and azimuthal angle in the Bloch sphere, respectively. (\romannumeral3) perform a unitary evolution on the initial state with a parameter-dependent Hamiltonian $H\left(\lambda\right)$. (\romannumeral4) measure the observable of the final state $\left|\psi_{f}\right\rangle $ and perform data estimation. To establish the connection between the classical trajectories and the parameter precision in step (\romannumeral2), we introduce the quantum Cramér-Rao bound, which gives the lower bound on the variance of parameter $\lambda$, i.e., $\delta^{2}\lambda\geqslant \left(\nu\mathcal{F}_{q}\right)^{-1}$, where $\nu$ is the number of independent measurement rounds and $\mathcal{F}_{q} = 4\left\langle \partial_{\lambda}\psi_{f}|\partial_{\lambda}\psi_{f}\right\rangle -4\left|\left\langle \psi_{f}|\partial_{\lambda}\psi_{f}\right\rangle \right|^{2}$ is the QFI~\cite{RevModPhys.90.035005,liu2020quantum,doi:10.1142/S0217979210056335}. Then we prepare spin coherent states at each phase-space point, perform a unitary evolution under the Hamiltonian $H\left(\lambda\right)$, and ultimately obtain the distribution of the QFI across the entire phase space. By correlating the geometric structure of phase-space portrait with the QFI distribution, we can identify the regions where spin coherent states are initially located that lead to optimal metrological performance. 
	
	\textit{Two-mode BEC model}.--- Now we consider a specific example. We begin with the well-known two-mode Hamiltonian, which is
	  described as follows~\cite{RevModPhys.71.463,RevModPhys.73.307,PhysRevA.57.1208,PhysRevA.57.2920,PhysRevA.55.4318,PhysRevA.67.013607,PhysRevA.74.063614}

	\begin{equation}\label{eq1}
		H=\Omega S_{x}-\chi S_{z}^{2},
	\end{equation}
	 where $S_{x}=\left(a^{\dagger}b+b^{\dagger}a\right)/2$, $S_{y}=\left(a^{\dagger}b-b^{\dagger}a\right)/2i$ and $S_{z}=\left(a^{\dagger}a-b^{\dagger}b\right)/2$ are the collective spin operators. Here $a^{\dagger}$ ($a$) and $b^{\dagger}$ ($b$) are bosonic creation (annihilation) operators of two modes. The parameter $\Omega$ denotes the effective Rabi frequency which describes the coupling strength between the two modes, and $\chi$ denotes the effective interaction strength between the condensate particles. The above two-mode Hamiltonian can be realized in a system with a double-well trapping potential~\cite{PhysRevA.55.4318,Sørensen2001,PhysRevA.67.013607,PhysRevA.72.033612,PhysRevA.71.043603,PhysRevA.74.063614}. To analyze the dynamics, we employ a semiclassical approach to compute the phase-space trajectories corresponding to Eq.~\eqref{eq1}.  In thermodynamic limit of the total particle number $N\rightarrow\infty$, the quantum fluctuations can be neglected,  enabling us to replace the creation and annihilation operators with $c$ numbers, i.e., $a\rightarrow\sqrt{N_{a}}e^{i\phi_{a}}$ and  $b\rightarrow\sqrt{N_{b}}e^{i\phi_{b}}$, where $N_{a}+N_{b}=N$. Applying the above relationship, the collective spin operators becomes
	 $S_{x}\rightarrow\frac{N}{2}\sqrt{1-z^{2}}\cos\phi$, $S_{y}\rightarrow\frac{N}{2}\sqrt{1-z^{2}}\sin\phi$, and $S_{z}\rightarrow\frac{N}{2}z$. The mean-field energy of Eq.~\eqref{eq1} is written as 
	 \begin{equation}\label{eq2}
	 	H_{\rm MF}=\frac{N\Omega}{2}\sqrt{1-z^{2}}\cos\phi-\frac{N^2\chi}{4}z^{2},
	 \end{equation}
	  where $\phi=\phi_{b}-\phi_{a}$ is the relative phase and  $z=\cos\theta=\left(N_{a}-N_{b}\right)/N$ is the population difference. Utilizing Eq.~\eqref{eq2} we have the equations of motion for canonical variables $\phi$ and $z$ which are 
	  \begin{align}
	  	\frac{d\phi}{dt}&=\frac{2}{N}\frac{\partial H_{\rm MF}}{\partial z}=-N\chi z-\frac{\Omega z\cos\phi}{\sqrt{1-z^{2}}}=f\left(\phi,z\right),\label{eq3}\\\frac{dz}{dt}&=-\frac{2}{N}\frac{\partial H_{\rm MF}}{\partial\phi}=\Omega\sqrt{1-z^{2}}\sin\phi=g\left(\phi,z\right). \label{eq4}
	  \end{align}
	  
\begin{figure}[tbp]
		\includegraphics[width=\linewidth]{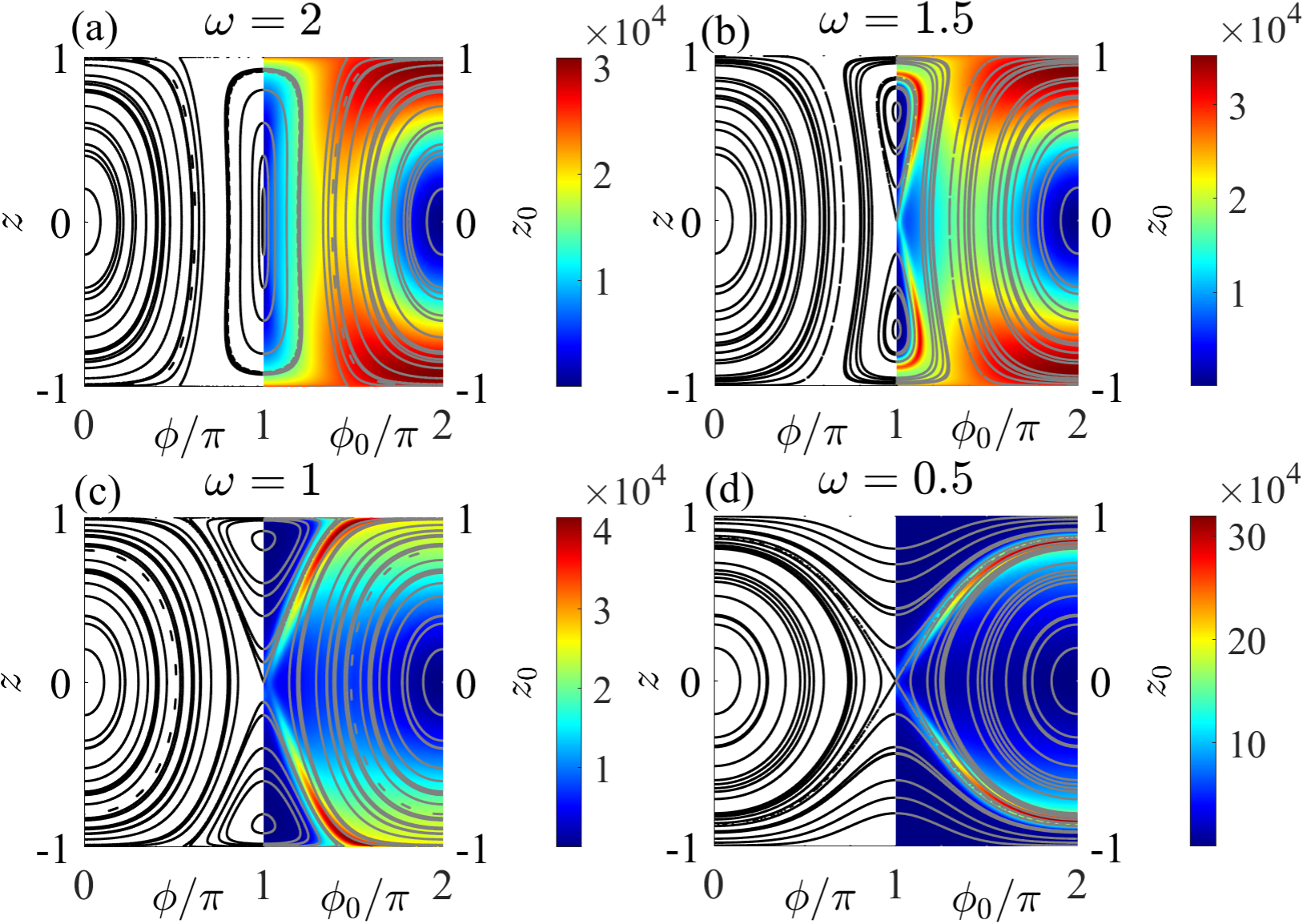}
		\caption{\label{fig:2} (a)-(d)  Left side of each panel: The phase-space portrait of the classical motion trajectories of the two-mode Hamiltonian with different parameters $\omega=2$, $1.5$, $1$ and $0.5$, respectively. Similar to a pendulum, only oscillatory modes exist in the strong-coupling regime ($\omega > 1$). In the weak-coupling regime ($\omega < 1$), rotational modes emerge, corresponding to the trajectories distributed above the upper and below the lower separatrices. Right side of each panel: The distribution of the corresponding QFI in phase space. The other parameters chosen here are $\chi=0.04$, $N=1000$ and $t=200\left(N\chi\right)^{-1}$.
		}
	\end{figure}

	By numerically solving the Eqs.~\eqref{eq3} and \eqref{eq4}, we obtain the phase-space trajectories under the mean-field approximation, as shown in Fig.~\ref{fig:2}. Since the trajectory is symmetric about $\phi=\pi$, we focus on the left half panel. For convenience, we define a dimensionless parameter $\omega\equiv2\Omega/\chi N$ to distinguish between different coupling strengths.  In Figs.~\ref{fig:2}~(a)-(d), we give the trajectories in phase space with $\omega=0.5$, $1$, $1.5$ and $2$.  The fixed points make the functions in Eqs.~\eqref{eq3} and \eqref{eq4}  satisfy  $f\left(\phi,z\right)=g\left(\phi,z\right)=0$ are  $\left(\phi,z\right)=\left(0,0\right)$, $\left(\pi,0\right)$  and $\left(2\pi,0\right)$, respectively. The properties of the fixed points are determined by the eigenvalues of the Jacobian matrix of the functions  $f\left(\phi,z\right)$ and $g\left(\phi,z\right)$  with respect to  $\left(\phi,z\right)$~\footnote{The Jacobian matrix for the variables $\bm{R}=\left(\phi,z\right)$ is defined as  $\mathcal{J}=\begin{pmatrix}
			\partial f/\partial\phi & \partial f/\partial z \\
			\partial g/\partial\phi & \partial g/\partial z
		\end{pmatrix}=\begin{pmatrix}
		\frac{\Omega z\sin\phi}{\sqrt{1-z^{2}}} & -N\chi-\frac{\Omega\cos\phi}{\left(1-z^{2}\right)^{3/2}}\\
		\Omega\sqrt{1-z^{2}}\cos\phi & -\frac{\Omega z\sin\phi}{\sqrt{1-z^{2}}}
		\end{pmatrix}$}.  
	For the fixed point $\left(\pi,0\right)$, the eigenvalues are $\pm\sqrt{\left(N\chi-\Omega\right)\Omega}$, thus the point $\left(\pi,0\right)$ is a saddle point when $\omega<2$, whereas it becomes an elliptic point or stable point when $\omega>2$. For the other fixed points, imaginary eigenvalues indicate they are stable points.

	The corresponding results are depicted on the right side of Figs.~\ref{fig:2}~(a)-(d). A significant correlation can be clearly seen between the distribution of the phase-space portrait (left side of each panel) and the QFI (right side of each panel).  The primary distinction across values of 
	$\omega$ lies in the variation of the phase-space structure around the fixed point  $(\pi,0)$. In the strong coupling regime ($\omega>1$), the trajectories shown in Figs.~\ref{fig:2}~(a) and (b) solely exhibit oscillatory modes~\cite{PhysRevA.67.013607}. The right side of Figs.~\ref{fig:2}~(a) and (b) shows that the regions with larger QFI appear along the separatrices between the different oscillatory modes. As $\omega$ increases to $2$, the previously separated oscillatory modes coalesce into one, and  $(\pi,0)$  becomes a stable point. In the weak-coupling regime ($\omega < 1$), the rotational modes emerge and the separatrices passing through the saddle point $(\pi, 0)$ separate the rotational modes from the oscillatory modes.

	 In Figs.~\ref{fig:2}~(a)-(d), we fix the evolution time 
	 $t$ and interaction strength $\chi$. While the Rabi frequency $\Omega$ decreases with $\omega$,  the maximal value of QFI increases and becomes localized along the separatrices. Specifically, as shown on the right side of Figs.~\ref{fig:2}~(c) and (d), larger values of the QFI are concentrated along the separatrices between the rotational and oscillatory modes (the red regions in Fig.~\ref{fig:2}~(c) with $\mathcal{F}_{q}\sim 4\times10^{4}$  and Fig.~\ref{fig:2}~(d) with $\mathcal{F}_{q}\sim 3\times10^{5}$). The dynamics along the separatrices, particularly those distinguishing between the oscillatory modes and the rotational modes, are more sensitive to parameter variation. Through the analysis based on the classical-quantum correspondence, the quantum states within these regions are more metrologically useful. For unstable points such as saddle points in classical dynamics, deviations in initial conditions cause the motion to depart from its original trajectory, which can be leveraged for quantum metrology. It is important to emphasize that the points on the separatrices exhibit performance comparable to or even better than that of the saddle points, which has not been previously reported.  We also calculate the scaling of QFI with particle number $N$ and find that the Heisenberg scaling $\mathcal{F}_{q}\sim N^2$ can be achieved at the separatrix, while it is not achieved in other areas~\cite{footnote2}. 
	 
	  In criticality-enhanced protocols, the QFI is related to the energy gap $\sqrt{\Delta}$, i.e., $\mathcal{F}_{q}\sim \Delta^{-3}$~\cite{PhysRevLett.126.010502}. As the control parameter approaches the critical point, the closure of the energy gap leads to an increase in the QFI, and the corresponding precision can attain the Heisenberg scaling~\cite{PhysRevLett.126.010502,PhysRevA.110.012413,Zhu2023}. However, maintaining the parameter around the critical point with the aforementioned method may increase difficulty in practical experiments, thus an alternative approach is necessary.

\begin{figure}[tbp]
	\includegraphics[width=\linewidth]{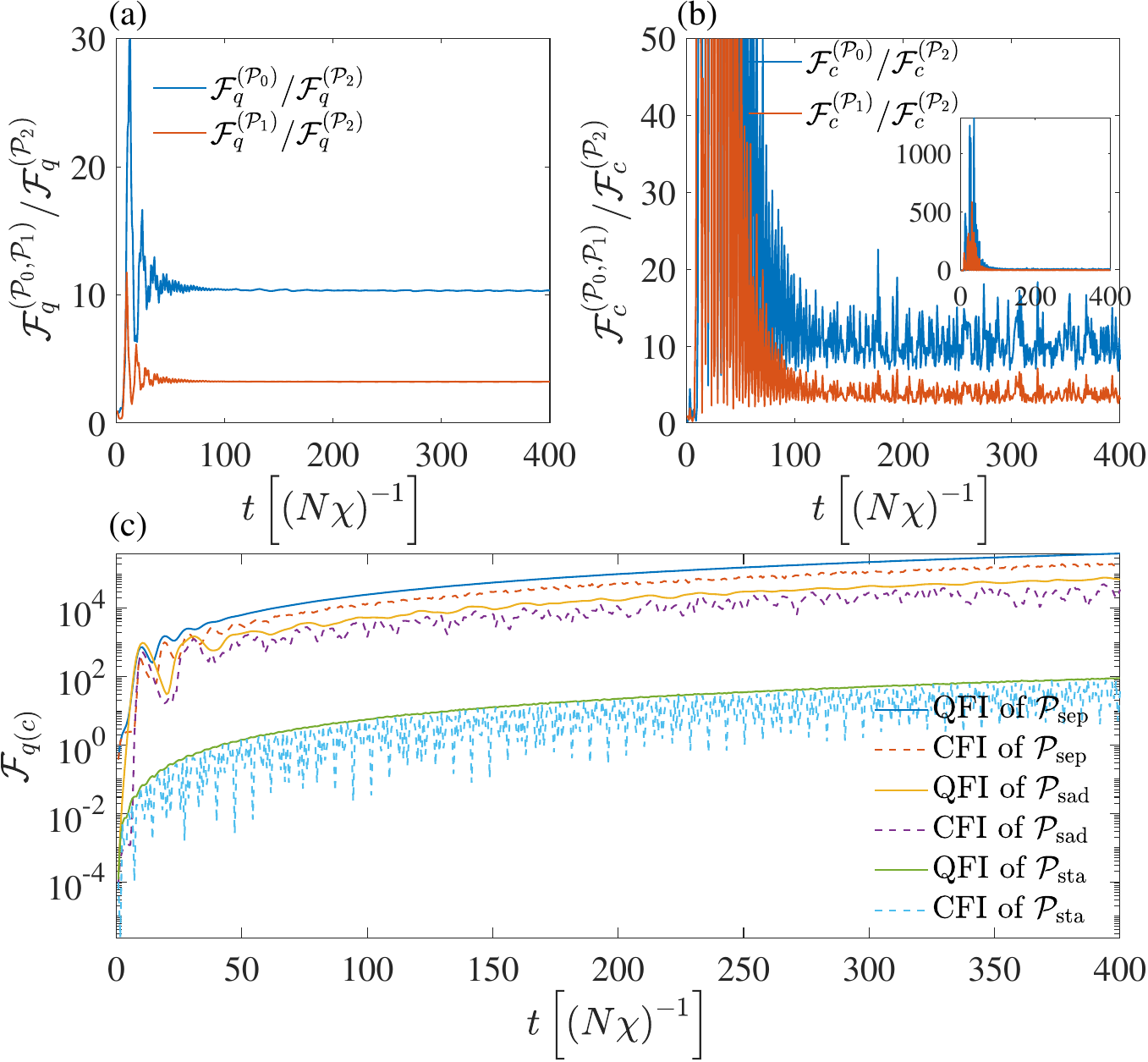}
	\caption{\label{fig:3}  (a) Ratios as functions of time of the QFI at its maximum for $\omega=0.5$ and $\omega=1$ to that for the critical case ($\omega=2$). (b) Ratios as functions of time of the CFI at its maximum for $\omega=1$ and $\omega=0.5$ to that for the critical case. The inset shows the ratio $\mathcal{F}_{c}^{(\mathcal{P}_{0},\mathcal{P}_{1})} / \mathcal{F}_{c}^{(\mathcal{P}_{2})}$ over a wider range of sizes. (c) QFI and CFI as functions of evolution time. From top to bottom, the three solid lines represent the QFI for initial states located at the point on the separatrix $\mathcal{P}_{\rm sep}$,  the saddle point $\mathcal{P}_{\rm sad}$, and the stable point $\mathcal{P}_{\rm sta}$, respectively; Similarly, the three dashed lines denote the corresponding CFI for these three cases.}
\end{figure}

	  Turning to the two-mode Hamiltonian in Eq.~\eqref{eq1}, the critical point is $\omega=\omega_{c}=2$, which corresponds to the case shown in Fig.~\ref{fig:2}~(a). The approximate analytical result of the QFI for this model is~\cite{PhysRevLett.126.010502,PhysRevA.110.012413}
	 \begin{equation}\label{eq6}
	 	\mathcal{F}_{q}\simeq\frac{\left[\sin\left(2\sqrt{\Delta}t\right)-2\sqrt{\Delta}t\right]^{2}}{16\Delta^{-3}}{\rm {\rm Var}}\left(D\right)_{\left|\Psi\right\rangle }
	 \end{equation}
	 where $\operatorname{Var}(\cdot)$ denotes the variance, $\Delta = \Omega(\Omega - N\chi)$, and $D = (\Omega - N\chi)\Omega x^{2} - \Omega p^{2}$ with $x \simeq \sqrt{2/N}S_{x}$ and $p \simeq -\sqrt{2/N}S_{y}$ defined by the Holstein-Primakoff transformation. However, this approximation holds only in the short-time regime. In the End Matter, we present an approximate analytical expression for the QFI under long-time evolution and elucidates the enhancement mechanism of the saddle point and separatrix. To facilitate a better comparison, we select the positions $\mathcal{P}_{0}$, $\mathcal{P}_{1}$, and $\mathcal{P}_{2}$ where the QFI is maximal in Figs.~\ref{fig:2}~(a), (c) and (d), and calculate the time evolution of the QFI for initial states located at these positions. Note that the maxima in Figs.~\ref{fig:2}~(c) and (d) are both located on the separatrices. In Fig.~\ref{fig:3}~(a), we present the ratios of the QFI for $\omega=0.5$ and $\omega=1$ to that of the critical case ($\omega=2$) as a function of time, i.e., $\mathcal{F}_{q}^{\left(\mathcal{P}_{0}\right)}/\mathcal{F}_{q}^{\left(\mathcal{P}_{2}\right)}$ and $\mathcal{F}_{q}^{\left(\mathcal{P}_{1}\right)}/\mathcal{F}_{q}^{\left(\mathcal{P}_{2}\right)}$, respectively.  In the long-time limit, we have $\mathcal{F}_{q}^{\left(\mathcal{P}_{0}\right)}/\mathcal{F}_{q}^{\left(\mathcal{P}_{2}\right)} \simeq 10$ and $\mathcal{F}_{q}^{\left(\mathcal{P}_{1}\right)}/\mathcal{F}_{q}^{\left(\mathcal{P}_{2}\right)} \simeq 3$.  For practical measurements, we evaluate the precision using the CFI. Its ratios, shown in Fig.~\ref{fig:3}~(b), display more pronounced oscillations than the QFI~\cite{PhysRevResearch.5.013185,footnote2}. Figure~\ref{fig:3}~(c) presents a comparison of the QFI and CFI at different locations, showing that exploiting the separatrix for metrology still provides a significant advantage. Furthermore, even accounting for decoherence in realistic systems, initial states on the separatrix continue to exhibit superior performance~\cite{footnote2,PhysRevLett.101.200402,PhysRevA.92.023603,PhysRevLett.105.020401,PhysRevA.110.032612,PhysRevLett.105.020401}. The above results indicate that the QSI scheme outperforms the criticality-enhanced scheme. Moreover, the QSI scheme shows lower parameter dependence (e.g., requiring only $\omega \le 1$ for two-mode BECs),  which enhances experimental feasibility.
	\begin{figure}[tbp]
	\includegraphics[width=\linewidth]{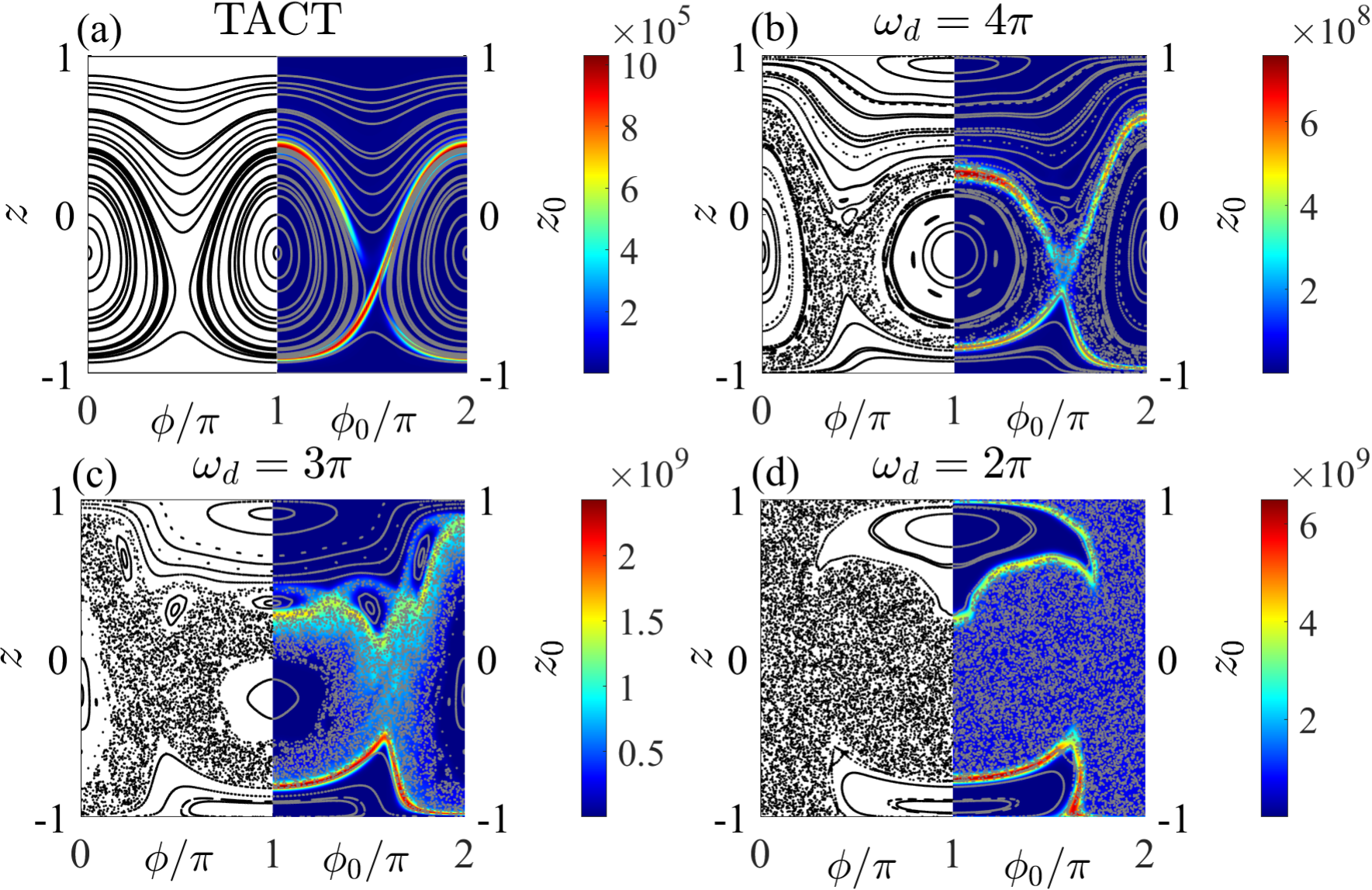}
	\caption{\label{fig:4} (a) Left side of the panel: The phase-space portrait of the classical motion trajectories of the TACT Hamiltonian. Right side of the panel: The distribution of the QFI for the TACT Hamiltonian in phase space.  (b)-(d) Left side of each panel: Poincaré sections for driving frequencies $\omega_{d} = 4\pi$, $3\pi$, and $2\pi$ based on Eq.~\eqref{eq9}. Right side of each panel: The distribution of the QFI for driving frequencies $\omega_{d} = 4\pi$, $3\pi$, and $2\pi$. }
	\end{figure}
	
	\textit{Cases of two-axis countertwisting model and chaotic dynamics}.--- Now we consider the Hamiltonian including the two-axis countertwisting (TACT)  interaction, which is also frequently used in nonlinear dynamics and expressed as 
	\begin{equation}\label{eq8}
	H_{\rm TACT}=\Omega' S_{z}+\chi'\left(S_{z}^{2}-S_{x}^{2}\right),
	\end{equation}  
	where $\Omega'$ denotes the effective transition frequency induced by an external field, and $\chi'$ denotes the strength of the interaction.  The preparation of highly spin-squeezed states using TACT interactions for parameter estimation has been discussed in Refs.~\cite{PhysRevA.90.062132,PhysRevA.92.013623,PRXQuantum.4.020314}, and here we focus more on directly utilizing its dynamics for parameter estimation. Similarly, the classical equations of motion are  $d\phi/dt=\Omega'+\chi'z\left(1+\cos^{2}\phi\right)$ and $dz/dt=-\frac{\chi'}{2}\left(1-z^{2}\right)\sin\left(2\phi\right)$. After a straightforward calculation, we obtain the corresponding saddle points in the phase space are  $\left(\phi,z\right)=\left(\pi/2,-\Omega'/N\chi'\right)$ and $\left(3\pi/2,-\Omega'/N\chi'\right)$ when $N\chi'>\Omega'$. In Fig.~\ref{fig:4}~(a), we obtain the phase-space portrait of the TACT Hamiltonian (left side) and the corresponding distribution of the QFI with respect to the parameter $\Omega'$ (right side). It is clear that the QFI along the separatrices, located between the oscillation and rotation modes and passing through the saddle points, remains higher than other regions, consistent with the conclusions drawn in the previous section.
	
	Chaotic dynamics can be induced by applying a periodic driving to the two-mode BEC.  Specifically, the Hamiltonian of this system is given by
	\begin{equation}\label{eq9}
	H=\Omega_{z}S_{z}+\chi_{z}S_{z}^{2}+\Omega_{x}\cos\left(\omega_{d} t\right)S_{x},
	\end{equation}
	  where the third term represents an external driving with the modulation frequency $\omega_{d}$. The results in Ref.~\cite{PhysRevA.103.023309} demonstrate that,  with the appropriate nonlinear strength $\chi_{z}$ and modulation frequency $\omega_{d}$, the system undergoes chaotic dynamics, which can facilitate the generation of entanglement beneficial for quantum metrology. In fact, we can induce a transition from the TACT dynamics to the chaotic dynamics by adjusting the frequency  $\omega_{d}$~\cite{footnote2,PhysRevA.103.023309,PhysRevA.91.043642,PhysRevA.107.042613}. The left side of Figs.~\ref{fig:4}~(a)-(d) displays the Poincaré sections for different frequencies, which are obtained by numerically calculating the phase-space positions for each period $T=2\pi/\omega_{d}$. Meanwhile, on the right side of Figs.~\ref{fig:4}~(a)-(d), we also exhibit the distribution of the QFI after evolving for $300 T$ from the initial spin coherent state $\left|\theta_{0},\phi_{0}\right\rangle$ towards various points in phase space. It shows that the chaotic regions spread out from the saddle points and grow more extensive as the frequency decreases. Even though the oscillatory and rotational modes have disappeared, the separatrices between the chaotic and regular regions still exhibit higher QFI.

	 \textit{Conclusion}.---
	To summarize, we propose a QSI scheme that exploits nonlinear dynamics for parameter encoding, thereby enhancing measurement precision.   By employing the quantum-classical correspondence and selecting the saddle points along with the separatrices passing through them, we obtain a larger QFI compared to other regions. In contrast to criticality-enhanced protocols that rely on vanishing energy gaps to improve precision, the approach based on nonlinear dynamics provides a broader range of applicable parameters, and exhibits better feasibility in practical physical systems.
	Our scheme is applicable not only to models with one-axis interactions, such as the two-mode BEC model, but also to models with TACT interactions and those exhibiting chaotic dynamics. The present work provides insights that could inform the design of experimentally feasible quantum precision measurement schemes.

	\begin{acknowledgments}
		We acknowledge Dr.~Zhucheng Zhang, Dr.~Chong Ye and Dr.~Wenkui Ding for valuable suggestions. This work was supported by the National Natural Science Foundation of China (Grants No.~12088101, No.~U2330401 and No.~12347127) and the Fundamental Research Funds for the Central Universities (Grant No.~2412023QD007).
	\end{acknowledgments}

\bibliography{mybib}

	\appendix

\onecolumngrid 
\vspace{24pt} 

\begin{center}
	\textbf{\large End Matter} 
\end{center}

\begin{figure}[tpb]
	\includegraphics[width=\linewidth]{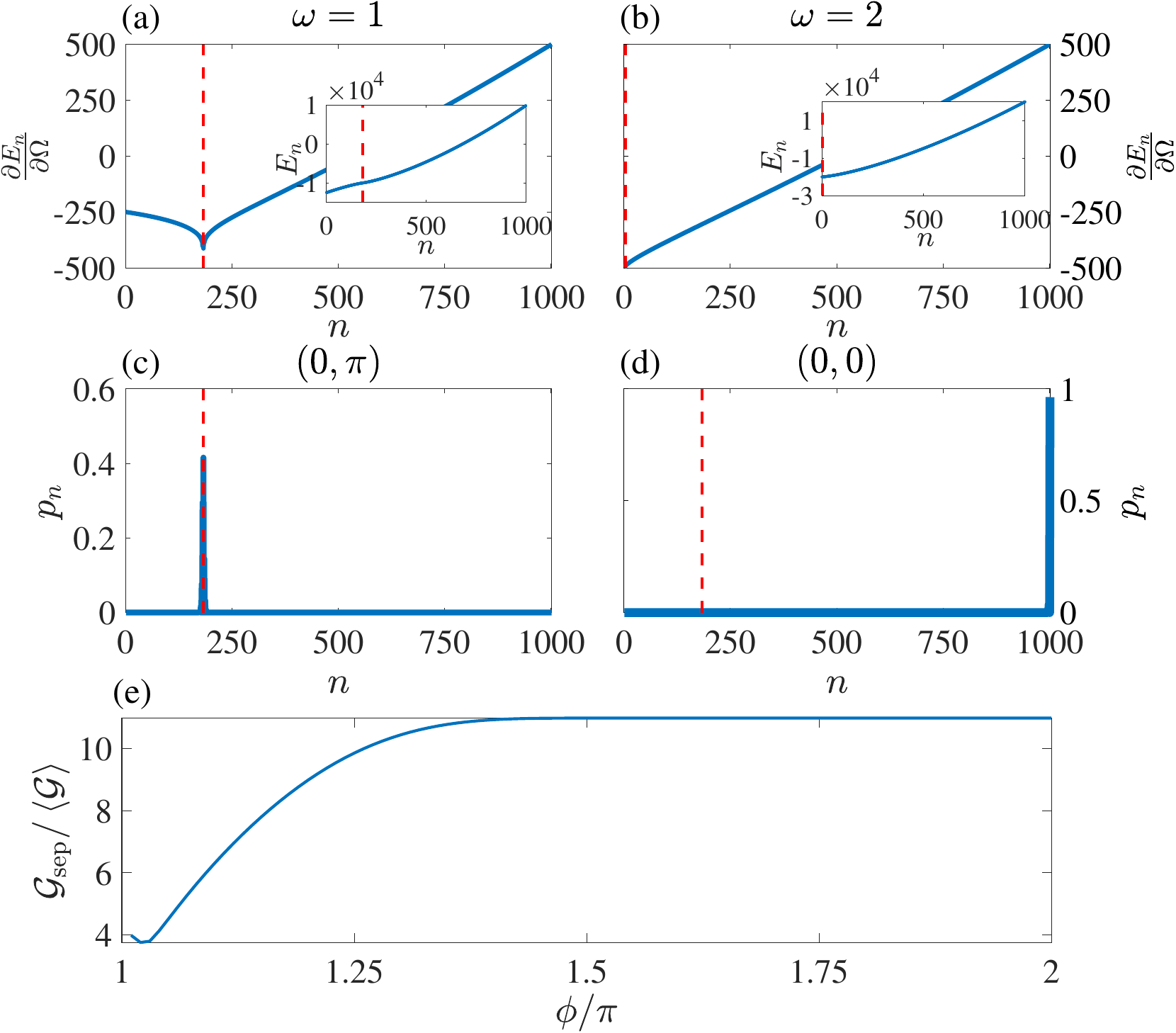}
	\caption{\label{fig:5} (a)-(b) Plots of the derivatives of energy spectrum with respect to the parameter $\Omega$ for two-mode Hamiltonian with  $\omega=1$ and $\omega=2$, respectively. In (a), the red line indicates the position of the minimum. The inset shows the transition of energy levels from being doubly degenerate to non-degenerate. The position of the minimum shifts leftward as $\omega$ increases, approaching $n=0$ when $\omega=2$, as shown in (b). (c)-(d) Population distribution $p_{n}$ for the saddle point $\left(0,\pi\right)$ and $\left(0,0\right)$ when $\omega=1$. For the saddle point, its distribution is centered around the minimum, whereas that of the stable point is located far from it. (e) Ratio of the gain factor $\mathcal{G}_{\rm sep}$ of the points on the separatrices within the range $\left[\pi,2\pi\right]$ to the average gain factor of $\left\langle \mathcal{G}\right\rangle$ at other positions in the phase space (excluding the separatrices). }
\end{figure}

\vspace{12pt} 
\twocolumngrid
	\textit{Enhancement Mechanism}.--- We further explore how the saddle point and separatrices enhance precision. Notably, the approximation in Eq.~\eqref{eq6} is confined to the low-excitation regime and only effectively characterizing the short-time evolution of the QFI, while the long-time QFI still lacks an effective description in collective spin systems. For a general initial state $|\psi_{\text{in}}\rangle = \sum_{n=0}^{N} c_n |\psi_n\rangle$, where $c_n = \langle \psi_n | \psi_{\text{in}}\rangle$ are the expansion coefficients in the eigenstates of the Hamiltonian, the QFI is given by

\begin{equation}\label{eq7}
	\mathcal{F}_{q}\simeq4\mathrm{ Tr}\left(\mathcal{E}^{2}\mathcal{C}\right)t^{2}-4\mathrm{ Tr}^{2}\left(\mathcal{C}\mathcal{E}\right)t^{2},
\end{equation} 
where $t$ is the evolution time. The symbols  $\mathcal{C}$ and $\mathcal{E}$ denote $(N+1)$-dimensional matrices with elements given by $\mathcal{C}={\rm diag}\left(\left|c_{0}\right|^{2},\left|c_{1}\right|^{2},\ldots,\left|c_{N}\right|^{2}\right)$ and $\mathcal{E}={\rm diag}\left(\partial_{\Omega}E_{0},\partial_{\Omega}E_{1},\ldots,\partial_{\Omega}E_{N}\right)$, respectively~\cite{footnote2}. From Eq.~\eqref{eq7}, the QFI is mathematically equivalent to the variance of $\partial E_{n}/\partial_{\Omega}$ weighted by the probability distribution $p_{n}=\left|c_{n}\right|^2$. Here we define the gain factor
\begin{equation}
 \mathcal{G}={\rm Tr}\left(\mathcal{E}^{2}\mathcal{C}\right)-{\rm Tr}^{2}\left(\mathcal{C}\mathcal{E}\right),
\end{equation}
whose value is primarily influenced by the parameter $\omega$ and the choice of the initial state.
Figures~\ref{fig:5}~(a)-(b) show the derivatives of the energy spectrum with respect to $\Omega$ for the two-mode Hamiltonian ($\omega=1, 2$), with energy levels labeled by integer $n$. For $\omega<2$, $\partial E_{n}/\partial{\Omega}$ exhibits a minimum at a certain energy level (indicated by the red lines). In the inset of Fig.~\ref{fig:5}~(a), the position of this minimum corresponds to the transition of the energy levels from a double degeneracy to non-degeneracy. The minimum shifts leftward as $\omega$ increases and vanishes when $\omega>2$.

For the gain factor $\mathcal{G}$, its maximization crucially hinges on harnessing this minimum.  Figure~\ref{fig:5}~(c) shows that when $\omega=1$,  the population distribution $p_{n}$ for the saddle point $\left(0,\pi\right)$ precisely centered at this minimum. In contrast, as shown in Fig.~\ref{fig:5}~(d), the distribution $p_{n}$ for the stable point $\left(0,0\right)$ deviates from the minimum. A straightforward calculation demonstrates that selecting the saddle point position as the initial state yields a larger gain factor $\mathcal{G}$. 

The population distribution for the points on the separatrices is also centered around the minimum~\cite{footnote2}. They are broader than the saddle-point distribution, further amplifying the gain factor $\mathcal{G}$.  Figure~\ref{fig:5}~(e) further illustrates the contrast between the gain factor $\mathcal{G}_{\rm sep}$ obtained along the separatrices within the interval $[\pi, 2\pi]$ and the value $\langle \mathcal{G} \rangle$ averaged across all other phase-space regions where the separatrices are excluded. It shows the gain factor on the separatrices $\mathcal{G}_{\rm sep}$ is superior to that at other positions, making the separatrices more suitable for selection as the initial states.

\end{document}